\documentclass[
]{article}
\date{}

\usepackage{authblk}
\usepackage{booktabs}
\usepackage{multirow}
\usepackage{graphicx}
\usepackage[numbers]{natbib}
\usepackage{listings}
\usepackage{lineno}
\title{Exploratory Models of Human-AI Teams: Leveraging Human Digital Twins to Investigate Trust Development}

\usepackage{authblk}   
\usepackage{hyperref}  
\usepackage{footmisc}  

\author[1]{Daniel Nguyen}
\author[1,2]{Myke C. Cohen}
\author[1]{Hsien-Te Kao}
\author[1]{Grant Engberson}
\author[1]{Louis Penafiel}
\author[1]{Spencer Lynch}
\author[1]{Svitlana Volkova}

\affil[1]{Aptima, Inc., Woburn, MA}
\affil[2]{Human Systems Engineering, Arizona State University, Mesa, AZ}

\begin{document}

\maketitle


\begin{abstract}
As human-agent teaming (HAT) research continues to grow, computational methods for modeling HAT behaviors and measuring HAT effectiveness also continue to develop. One rising method involves the use of human digital twins (HDT) to approximate human behaviors and socio-emotional-cognitive reactions to AI-driven agent team members (Barricelli \& Fogli, 2024). In this paper, we address three research questions relating to the use of digital twins for modeling trust in HATs. First, to address the question of how we can appropriately model and operationalize HAT trust through HDT HAT experiments, we conducted causal analytics of team communication data to understand the impact of empathy, socio-cognitive, and emotional constructs on trust formation. Additionally, we reflect on the current state of the HAT trust science to discuss characteristics of HAT trust that must be replicable by a HDT such as individual differences in trust tendencies (e.g., propensity to trust, Jessup et al., 2019), emergent trust patterns (e.g., trust violation and repair, Wildman et al., 2024), and appropriate measurement of these characteristics (e.g., growth modeling, Abramov et al., 2020). Second, to address the question of how valid measures of HDT trust are for approximating human trust in HATs, we discuss the properties of HDT trust: self-report measures, interaction-based measures, and compliance type behavioral measures. Additionally, we share results of preliminary simulations comparing different LLM models for generating HDT communications and analyze their ability to replicate human-like trust dynamics. Third, to address how HAT experimental manipulations will extend to human digital twin studies, we share experimental design focusing on propensity to trust for HDTs vs. transparency and competency-based trust for AI agents. 
\end{abstract}

Keywords: 
  Human-AI Teaming,
  Trust,
  Human Digital Twins


\maketitle

\section{Introduction}
The integration of artificial intelligence (AI) into operational environments has become increasingly vital across diverse domains, fundamentally transforming how humans and machines collaborate to achieve shared objectives. Over the past decade, Human-AI Teaming (HAT) research has emerged as a critical field, with scholars applying cognitive science principles to understand the complexities of these novel partnerships. This growing body of research reflects the urgency of understanding how humans and AI can work together effectively, safely, and productively.

As the HAT literature expands and AI capabilities advance, innovative methodologies for studying these interactions have emerged. One promising approach involves the use of human digital twins (HDTs) - computational models designed to replicate human responses and behaviors within HAT contexts. These HDTs can be configured to simulate both state-based responses (such as transient cognitive and affective reactions) and trait-based characteristics (including dispositional individual differences), offering a versatile alternative to traditional human-subjects research \cite{barricelliDigitalTwinsHumanComputer2024, nationalacademiesofsciencesengineeringandmedicineFoundationalResearchGaps2024}.

The advantages of HDTs in HAT research are significant. Beyond addressing the practical constraints of human-subjects studies, such as cost and recruitment challenges, HDTs provide unprecedented control over experimental variables and the ability to rapidly test multiple scenarios. This capability is particularly valuable for investigating complex team phenomena, with trust emerging as a critical area of focus. Trust - the willingness to be vulnerable to another agent's actions - serves as a fundamental determinant of HAT effectiveness and success.

However, the validity of using HDTs to study trust dynamics in HATs requires careful examination. This paper addresses three crucial questions regarding the implementation of HDTs in trust research: (1) How can we effectively model and measure HAT trust using HDT-based approaches? (2) What are the essential characteristics of HAT trust that must be operationalized in HDT trust models? (3) How do experimental manipulations from traditional HAT studies translate to HDT-based research?

By examining these questions, we aim to establish a framework for validating and implementing HDTs in HAT trust research, ultimately advancing our understanding of human-AI collaboration and trust development.
\section{Research Question 1: How can we model and measure HAT trust dynamics in HDTs?}

\subsection{Theoretical Considerations for Modeling HAT Trust in HDTATs}

\subsubsection{Evolution of Trust Definitions in Human-AI Contexts}
Trust has emerged as a multifaceted construct whose definition has evolved alongside technological advancement. In organizational contexts where HATs operate, Mayer et al. \cite{mayerIntegrativeModelOrganizational1995} established a foundational definition of trust as an individual's willingness to be vulnerable to another agent's actions during goal-oriented, risky situations. This conceptualization gained particular relevance in AI-driven systems, where users must navigate the inherent risks of relying on imperfect automation \cite{parasuramanHumansAutomationUse1997}. Building on this foundation, Lee and See \cite{leeTrustAutomationDesigning2004} specifically addressed trust in automation, defining it as an attitude regarding an automated agent's reliability in risk-characterized situations. While this definition remains influential in HAT research, it primarily reflects traditional human-automation relationships characterized by clear supervisory hierarchies \cite{textorExploringRelationshipEthics2022}. Modern AI systems, however, engage in more sophisticated interactions that mirror human teamwork dynamics, suggesting trust should be reconceptualized as a dynamic, relational construct \cite{chiouTrustingAutomationDesigning2023}.

\subsubsection{Trust Formation and Reliance in Human-AI Teams}
Lee and See's \cite{leeTrustAutomationDesigning2004} theoretical model suggests that trust in automation develops through people's understanding of three key elements: the system's intended functions, expected performance levels, and underlying processes. This framework suggests that optimal trust—and consequently, appropriate reliance—emerges when users' perceptions align with the AI's actual capabilities \cite{parasuramanHumansAutomationUse1997, chanceyTrustComplianceReliance2017}. This alignment becomes particularly crucial in safety-critical domains like search and rescue operations or disaster response, where HAT failures can have severe consequences \cite{gliksonHumanTrustArtificial2020a}. The wealth of empirical literature on trust in AI makes it an ideal construct for exploring the utility of HDTAT experimental paradigms.

However, recent meta-analyses challenge the assumed relationship between trust attitudes and behavioral reliance \cite{hancockMetaAnalysisFactorsAffecting2011, pattonRelationshipTrustDependence2024, schaeferMetaAnalysisFactorsInfluencing2016}. Users may maintain general trust in an AI system while choosing not to rely on it in specific instances, highlighting the distinction between attitudinal and behavioral trust. This observation aligns with the argument that trust in automation is indisputable only when actual reliance is observed \cite{meyerTrustRelianceCompliance2013}. Consequently, HDTAT experiments must model both perceptual trust, which encompasses attitudes and perceptions about an AI teammate's trustworthiness, and behavioral trust, which manifests through observable decisions reflecting trust through actions.

\subsubsection{Beyond Capability-Based Trust}
Contemporary research reveals that trust in HATs extends beyond perceptions of AI capabilities. Jessup et al. \cite{jessupMeasurementPropensityTrust2019a} emphasize the role of automation trust propensity—an individual's baseline tendency to trust automated systems. Hoff and Bashir \cite{hoffTrustAutomationIntegrating2015} characterize this as dispositional trust, shaped by demographic and personality factors including age, gender, culture, and individual traits. While these variables prove difficult to control in traditional laboratory studies, HDTATs offer unique opportunities to explore their effects systematically, particularly regarding emotional and moral dimensions of trust that remain understudied \cite{malleChapter1Multidimensional2021}.

\subsubsection{Trust as a Dynamic Social Construct}
The validation of HDTAT trust models must acknowledge trust's dynamic and social nature. Research has identified temporal patterns in human-automation trust, including the ``perfect automation schema'' where users initially overestimate unfamiliar automation capabilities \cite{dzindoletRoleTrustAutomation2003}, and a tendency to undervalue AI reliability following errors \cite{snowSatisficingArtificingEvolution2021}. These patterns align with Lee and See's \cite{leeTrustAutomationDesigning2004} model stating that people initially trust automation using faith-based assumptions that it is reliable and eventually progress to more situational assessments of its dependability.

In HAT contexts, trust dynamics increasingly resemble interpersonal relationships \cite{madhavanSimilaritiesDifferencesHuman2007}, functioning as a continuous activity with varying manifestations across different timescales \cite{chiouTrustingAutomationDesigning2023}. While demographic influences on dispositional trust remain relatively stable \cite{jessupMeasurementPropensityTrust2019a}, perceptual and behavioral trust fluctuate based on observed AI reliability \cite{yangQuantifyingTrustDynamics2023, devisserTheoryLongitudinalTrust2020}. Successful HDTAT models must capture these nuanced trust dynamics, reflecting the sophisticated measures developed through decades of HAT research.

\subsection{Methods for Measuring HAT Trust}
Kohn et al. \cite{kohnMeasurementTrustAutomation2021} identified three main techniques for measuring trust in automated agents, corresponding to how they are obtained from human trustors: 1) self-reported trust, 2) behavioral or observational trust, and 3) physiological trust. Note our HDTAT experiments investigate self-reported and behavioral trust measurement techniques. 

\subsubsection{Self-reported Trust}
Self-reported trust measures are gathered through questionnaires administered at some point during a HAT experiment and are the most widely used trust measure in the literature \cite{kohnMeasurementTrustAutomation2021}. Trust questionnaires are often developed as psychometric tests tested through experimental setups (e.g., \cite{jianFoundationsEmpiricallyDetermined2000}) or construct-based reconstructions of existing surveys (e.g., Chancey et al.'s \cite{chanceyTrustComplianceReliance2017} adaptation of Madsen and Gregor's Human-Computer Trust scale \cite{madsenMeasuringHumancomputerTrust2000}). As such, self-reported techniques typically measure dispositional trust when administered at the beginning of an experiment, and perceptual trust when administered after a period of interaction with an AI counterpart \cite{cohenAnthropomorphismModeratesRelationships2023, hoffTrustAutomationIntegrating2015}.

While widely used, self-report measures face some validity challenges, including potential biases in how people assess and report their own trust levels. Individuals may struggle to accurately introspect about their trust, or social desirability biases could influence their responses. Some validity issues also arise from the rigorous demands of psychometric test development and the highly contextual nature of HAT experimentation. First, self-reported questionnaires are often administered as a repeated measure of perceptual trust throughout an experiment; however, the frequency and length of repeated surveys risk internal validity problems linked to survey fatigue and response anchoring effects \cite[ch. 2]{shadishExperimentalQuasiexperimentalDesigns2001}. However, this particular issue may be less of a problem in HDTATs as HDTs may be more consistent in answering repeatedly administered measures. 

A more pressing validity issue is that many HAT experiments use extensively modified versions of validated trust questionnaires, resulting in context-specific offshoot scales that do not measure trust in the way that their parent scales define it as a construct \cite{razinConvergingMeasuresEmergent2023}. Many of these modified scales assume that survey respondents have similar internal definitions of trust and other trust-related terminology, which is problematic; an additional problem is that it is uncertain how to approximate individual differences in understanding survey questionnaires within HDTs. This could be due to the inherent difficulty LLMs have with quantitative reasoning \cite{mirzadehGSMSymbolicUnderstandingLimitations2024} or an issue of some of these sociological survey scales being underrepresented in the training data. Researchers have found success when asking HDTs to rate their experiences in terms of importance on a scale from 1 to 10 as a means of enhancing memory synthesis through RAG \cite{parkGenerativeAgentsInteractive2023} which could be emulated as a means of generating reasonable survey response data while avoiding issues of contextual invariance caused by more obscure response scales. This will cause a discontinuity between real survey results that rely on a Likert scale or a 1 to 5 scale for example, requiring HDT responses to be rescaled. But this is imperfect given that human survey responses are also known to be affected by the way in which survey scales are presented \cite{decastellarnauClassificationResponseScale2018, yanImpactQuestionScale2018}.

\subsubsection{Behavioral Trust}
Behavioral trust measures are gathered through observable human behaviors within a HAT context. Though also widely used, the unique contexts of individual HAT experiments can result in arbitrary definitions of observation-based techniques for measuring trust \cite{kohnMeasurementTrustAutomation2021}, requiring careful consideration when used in HDTAT trust studies.

Consider, for instance, compliance-based measures, which describe humans' adherence to AI recommendations or commands as an indicator of trust. The rate at which people follow an AI agent's suggestions in decision-making scenarios can serve as more objective indicators of trust relative to self-reported data. But depending on how researchers interpret decision-making with automation, compliance can be easily conflated with reliance (i.e., delegation of a decision to an AI counterpart without seeing its recommendation; \cite{meyerTrustRelianceCompliance2013}). People have different tendencies to erroneously comply and rely on automation—and, in many cases, these mistakes also have inequal consequences \cite{chanceyTrustComplianceReliance2017}. The distinctions between different forms of behavioral trust can be blurrier in abstract HDTAT experimental settings, posing a trade-off between the range of possible HDT-AI decision-making outcomes within them and the fidelity of modeling real human decision-making biases with automation.

The dynamic and observable nature of interactions is another reason why behavioral techniques are widely used measures trust. Rather than asking about trust directly, interaction patterns can be observed and processed into quantifiable measures that serve as proxies for trust. For example, highly stable human communication patterns regarding sensitive task information with an AI counterpart could indicate higher levels of trust \cite{demirExplorationTeammateTrust2021}. Changes in the frequency and depth of interactions may also serve as behavioral indicators of trust development over time. As such, many researchers suggest tracking human-AI interactions over time to map fluctuations in trust and intervene where necessary \cite{devisserTheoryLongitudinalTrust2020}. But changes in interactions may not necessarily reflect changes in trust; under increased workload, people may accept machine counterpart’s suggestions more frequently despite not necessarily trusting it more \cite{karpinskyAutomationTrustAttention2018} (cf., \cite{satoAutomationTrustIncreases2020}). In HDTATs, this vagueness is complicated by the difficulty to control for the impacts of time on trust dynamics; without human input, many current HDT models are restricted to ordinal approximations of temporal interaction sequences that shape HMT trust.

As previously noted, behavioral trust measures may offer conflicting narratives with self-reported trust measures about how people are trusting an automated counterpart (Hancock et al., 2011). It is important to consider whether such conflicts in HDTATs arise from validity issues, considering how trust measures are gathered within HDT settings.

\subsubsection{Emerging Techniques for Measuring Trust}
In addition to these established measures, we recommend exploring several other potential types of HAT trust measures. {\it Physiological measures} could provide insight into subconscious trust reactions by tracking metrics like heart rate variability, skin conductance, or eye movements during human-AI interactions. These may capture subtle trust signals that people are not consciously aware of. Multi-modal trust measures that combine data across self-report, behavioral, physiological, and other channels may provide a more holistic view of trust as a complex, multi-faceted construct. Integrating diverse measure types could help overcome limitations of any single approach. {\it Longitudinal trust measures} that track trust development over extended periods of human-AI collaboration are also needed. These could reveal how trust forms, stabilizes, or degrades across the life cycle of human-AI relationships. 

One promising emergent trust measure for HDTATs involves the use of linguistic analysis of human-AI communication, particularly word choice, sentiment, and other language patterns. Notably, aside from a few examples (e.g., \cite{cohenDynamicsTrustVerbal2021, liEstimatingTrustConversational2022, liItsNotOnly2024}), research is scarce on the semantic and emotional facets of language that are key to conveying trust \cite{duanCommunicationHumanAITeaming2023}. Whereas extant HAT trust research guides how HDT trust should be formed and measured, additional research is needed to unpack the language of trust into the constituent social, emotional, and cognitive lexicon used to convey trust. How humans describe and refer to AI agents in natural conversation may indicate their trust orientations.

\subsection{Empirical Investigations into HAT Trust using Exploratory Causal Analysis}
To guide our causal analysis, we rely on extant theories that elucidate how social, emotional, and cognitive processes (i.e., socio-emotional-cognitive constructs) unfold in HATs (e.g., \cite{barsadeRippleEffectEmotional2002, forgasMoodJudgmentAffect1995}). The interplay between socio-emotional-cognitive constructs and how they manifest in team member interactions can be well understood through the affective infusion model \cite{forgasMoodJudgmentAffect1995} and emotional contagion theory \cite{barsadeRippleEffectEmotional2002}. As team members interact with one another, the affective infusion model explains that human team members will experience various emotions as the actions of other team members may trigger emotional reactions and affectively charge one’s thoughts regarding the team and its current state/actions~\cite{forgasMoodJudgmentAffect1995}. Other team members cue in to these affective tones, and thus one team member’s emotions may subsequently manifest in other team members through the emotional contagion process \cite{barsadeRippleEffectEmotional2002}. This process creates a team’s affective tone, which has been shown to influence constructs that vital indicators of healthy team functioning such as trust \cite{collinsGroupAffectiveTone2013} and performance \cite{linEffectTeamAffective2017}. 

Taken in concert, these theories underscore the importance of identifying key socio-emotional-cognitive constructs encoded in team communications. To this end, we conducted exploratory causal analyses of team communication data using the ``Saturn+'' dataset from DARPA’s Artificial Social Intelligence for Successful Teams (ASIST) program \cite{huangArtificialSocialIntelligence2022} to identify the impact of three key construct groups on trust: empathy, socio-cognitive, and emotional constructs (see Table \ref{tab:causal-analysis-overview}). We share the results of our causal structure learning analytics applied to analyze these data, and how these indicators predict each of the four indicators of trust used in the ASIST study (e.g., trust that the AI improved performance, trust in the AI’s recommendation, trust in the AI’s explanation of its recommendation, and trust that the AI improved team coordination). 

\begin{table}
    \centering
    \begin{tabular}{p{0.2\linewidth} p{0.4\linewidth} p{0.3\linewidth}}
    \toprule
         \textbf{Category} & \textbf{Constructs} & \textbf{Construct Analytics} \\ 
    \midrule
         \multirow{2}{*}{Empathy} & Empathy Intent & EmpGPT-3 \\
          & Empathy Emotions & EmpGPT-3 \\
    \midrule
         \multirow{3}{*}{Socio-cognitive} & Connotations, Perspectives, Attitudes & Lexicon-based analytics \\
          & Moral Values (Harm, Fairness, Purity, Authority, Ingroup) & Lexicon-based analytics \\
          & Subjectivity & Lexicon-based analytics \\
    \midrule
         \multirow{3}{*}{Emotions} & Sentiment & Hugging Face Sentiment \\
          & Toxicity & Detoxify \\
          & Emotions & Bert-base-uncased trained for emotions \\
    \bottomrule
    \end{tabular}
    \caption{Overview of Exploratory Causal Analyses}
    \label{tab:causal-analysis-overview}
\end{table}

Our goal is to identify key the strength and directionality between these latent socio-emotional-cognitive constructs and key indicators of team functioning (to inform higher fidelity HDT development and improve HAT trust over time. 

Casual structure learning involves identifying and modeling the relationships between variables in a way that captures the underlying causal mechanisms~\cite{sharmaDoWhyEndEndLibrary2020}. In this context, the interpretation of weights on causal relationships is crucial, as they often represent the strength and direction of influence between variables. Positive weights indicate a direct positive relationship, where an increase in one variable leads to an increase in another, while negative weights suggest an inverse relationship. However, interpreting these weights requires careful consideration of confounding factors and the potential for bias~\cite{ananthConfoundingCausalityConfusion2017}. Properly understanding these weights not only enhances the reliability of causal inferences but also aids in making informed predictions and decisions based on the modeled relationships.    

\subsubsection{Causal Analysis Results: How Empathy Constructs Effect HAT Trust?}
\begin{figure}
    \centering
    \includegraphics[width=\linewidth]{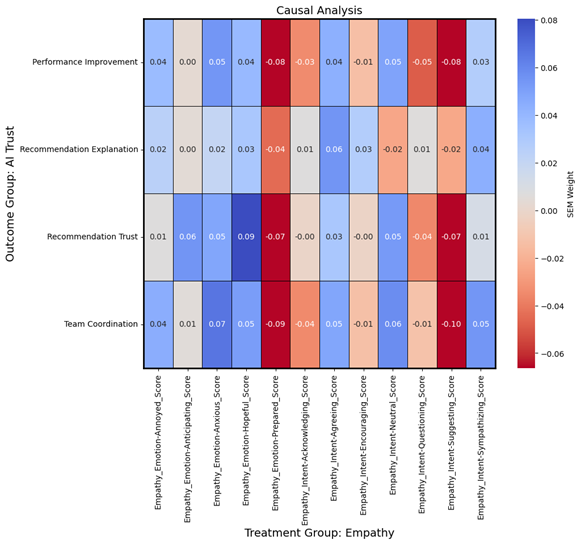}
    \caption{Results of causal analysis on how empathy constructs effect HAT trust.}
    \label{fig:causal-empathy}
\end{figure}

Empathy plays a critical role in fostering effective teamwork within collaborative environments and encompasses the intent to understand and the emotional resonance with team members. The ability to recognize and respond to the feelings and perspectives of others cultivates a supportive atmosphere, enhancing communication and trust among team members~\cite{leeTrustAutomationDesigning2004}. This emotional intelligence allows individuals to navigate conflicts with greater sensitivity, leading to more constructive resolutions~\cite{widayatiEffectEmotionalIntelligence2022}. Moreover, when team members demonstrate genuine empathy, they contribute to a collective sense of belonging and motivation, which can enhance group cohesion and performance~\cite{lucaDoesEmotionalIntelligence2001}. Within our analyses, we used the presence of specific empathetic strategies (e.g., agreeing, suggesting, and neutral responding) as indicators of empathy. Figure \ref{fig:causal-empathy} shows how empathy indicators predicted various indicators of trust in the AI.

The causal analysis of empathy revealed the relationships between various empathetic approaches and the outcomes related to AI trust and team coordination. Specifically, strategies focused on empathy intent—such as agreeing, neutral responses, and suggesting—demonstrated varying levels of effectiveness in fostering an environment conducive to collaboration. In particular, the findings indicate that empathetic communication can enhance trust in AI recommendations when deployed strategically, suggesting that the manner in which empathy is expressed can directly influence team dynamics and AI reliance.

Furthermore, the results suggest that while some empathetic responses promote positive outcomes such as team coordination, others may inadvertently lead to negative perceptions about trust and performance. For example, suggesting that empathetic responses appeared to cause declines in trust and team coordination. This paradox highlights the complexity of empathetic interactions, emphasizing that not all empathetic expressions yield beneficial results. Thus, the context and delivery of empathetic communication must be carefully considered to avoid potential pitfalls that could undermine team effectiveness.

Finally, the varying effects of emotional empathy, such as anticipatory and hopeful responses, underscore the importance of emotional intelligence in team settings. Emotional expressions that resonate with team members can increase trust in AI recommendations, thus improving overall performance and coordination. This analysis reinforces the notion that fostering an emotionally aware team culture, where empathetic communication is prioritized, can lead to improved AI trust and more effective teamwork. Ultimately, the nuanced interaction between empathy treatments and trust outcomes requires further exploration to refine communication strategies within collaborative environments.

\subsubsection{Causal Analysis Results: How Socio-cognitive Constructs Effect HAT Trust}

\begin{figure}
    \centering
    \includegraphics[width=\linewidth]{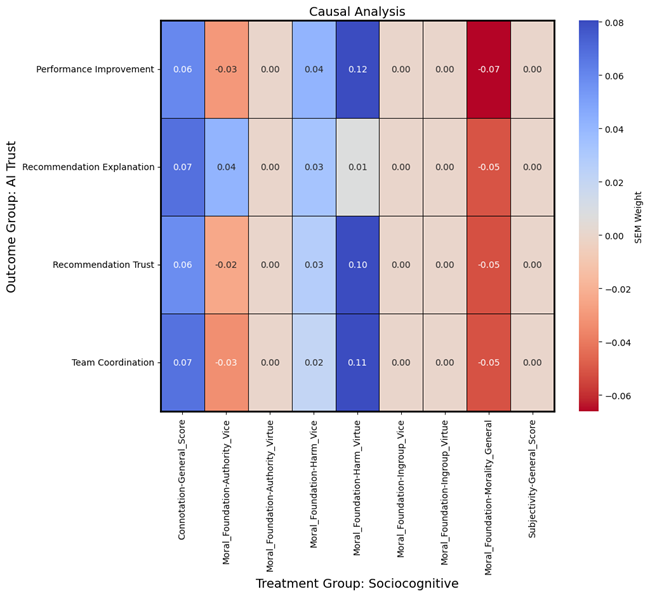}
    \caption{Results of exploratory causal analysis of socio-cognitive indicators on trust indicators.}
    \label{fig:causal-sociocog}
    \vspace{-0.5cm}
\end{figure}

The relationship between socio-cognitive elements such as subjectivity, connotation, and moral foundations, and teamwork in collaborative environments is crucial to promote effective communication and cohesion among team members. Subjectivity influences how individuals interpret shared information and experiences, shaping their perspectives and responses within the team context~\cite{rashkinConnotationFramesDataDriven2016,rashkinTruthVaryingShades2017}. Connotation further complicates this dynamic, as the emotional and cultural associations tied to language can lead to misunderstandings or reinforce biases that affect collaboration. Additionally, moral foundations underpin team values and ethical considerations, guiding decision-making and conflict resolution~\cite{gartenMoralityLinesDetecting2016}. In environments where these socio-cognitive factors are acknowledged and managed, teams are better equipped to leverage diverse viewpoints, enhance mutual respect, and achieve common goals, ultimately driving organizational success. Figure \ref{fig:causal-sociocog} depicts how indicators of connotation, moral foundation, and subjectivity predict various indicators of trust. 

The causal analysis of socio-cognitive treatments reveals that certain interventions, particularly those centered around connotation and moral foundations, can influence key outcomes related to team performance and AI trust. The findings indicate that fostering a positive connotation within team interactions is associated with enhancements in performance improvement and team coordination. This suggests that the way team members frame their communication can create a more conducive environment for collaboration, leading to better overall outcomes.

Conversely, the exploration of moral foundations highlights a more nuanced impact on team dynamics. While treatments focused on the harm virtue exhibit positive effects on team coordination and trust in recommendations, those centered on general morality may lead to adverse outcomes. This indicates that not all moral framing is equally effective; in some cases, it may even hinder trust and communication within the team. This complexity underscores the importance of selecting appropriate moral cues when designing interventions aimed at improving team interactions.

Finally, the relationship between these socio-cognitive treatments and AI trust outcomes is particularly noteworthy. The analysis reveals that effective communication strategies can bolster trust in AI recommendations, thereby enhancing overall team performance. However, the detrimental effects observed with certain moral framing approaches suggest that careful consideration is necessary when integrating ethical dimensions into team communication. This insight emphasizes the critical role that both connotation and moral foundations play in shaping trust dynamics within teams utilizing AI, highlighting the need for tailored strategies that align with the specific goals of team interactions.

\subsubsection{Causal Analysis Results: How Emotion Constructs Effect HAT Trust}
In teaming environments, the interplay between emotion—encompassing toxicity, sentiment, and specific emotional responses—plays a critical role in shaping team dynamics and overall effectiveness. Positive emotions, such as trust and enthusiasm, foster collaboration and innovation, encouraging open communication and shared problem-solving \cite{savaniDistilBERTEmotionRecognition2024}. Conversely, negative emotions, particularly toxicity manifested through conflict or resentment, can undermine cohesion and hinder performance \cite{hanuHowAILearning2021}. Understanding and managing these emotional undercurrents is essential for leaders and team members alike, as they directly influence decision-making, motivation, and interpersonal relationships \cite{sanhDistilBERTDistilledVersion2020}. Acknowledging the emotional landscape of a team not only enhances individual well-being but also cultivates a more resilient and productive collaborative culture. 

\begin{figure}
    \centering
    \includegraphics[width=\linewidth]{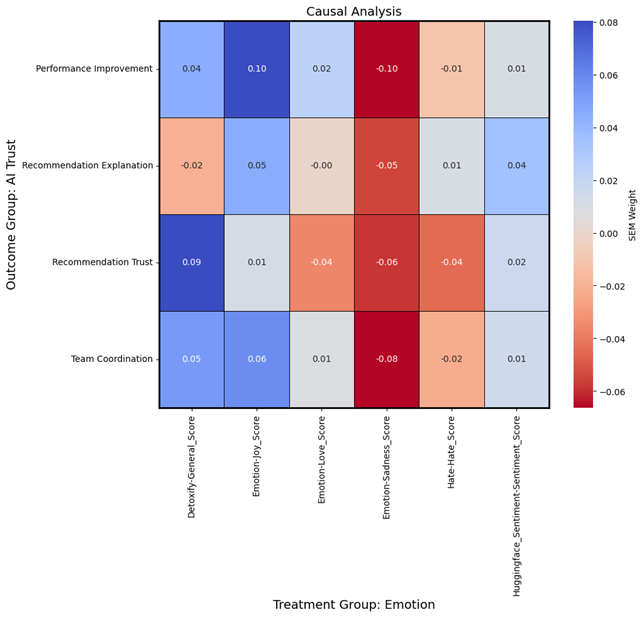}
    \caption{Results of exploratory causal analysis of emotion indicators on trust indicators.}
    \label{fig:causal-emotion}
    \vspace{-0.5cm}
\end{figure}

Within our emotion-construct analyses, we examine sentiment, toxicity, and discrete emotions. Figure \ref{fig:causal-emotion} depicts how indicators of emotion constructs predicted various indicators of trust in the AI.

The causal analysis reveals distinct patterns in the impact of emotion treatments on AI trust outcomes. Notably, interventions focused on promoting joy within the team setting are associated with improvements in performance and coordination. This suggests that fostering positive emotional states can enhance collaborative dynamics, leading to increased overall effectiveness. Teams that experience higher levels of joy are likely to engage more constructively, reinforcing the value of emotional well-being in achieving collective goals.

Conversely, the emotional treatment centered on sadness appears to have detrimental effects on performance improvement and team coordination. The findings indicate that negative emotional states can undermine communication, leading to decreased trust in AI recommendations. This highlights the importance of emotional climate within teams, as negative emotions can create barriers to effective collaboration and impede trust-building processes. The adverse impact of sadness underscores the need for proactive management of team emotions to maintain high levels of engagement and performance.

Lastly, the analysis identifies a nuanced relationship between emotional states and AI trust. While joy enhances trust in recommendations, sadness erodes it. This suggests that the emotional context in which teams operate plays a critical role in how AI outputs are perceived and accepted. Ensuring a positive emotional environment may not only facilitate better communication and coordination but also bolster AI trust, ultimately leading to more effective teamwork and decision-making. The findings emphasize the need for leaders to cultivate a supportive emotional atmosphere to maximize both human collaboration and AI reliance.

\subsubsection{Causal Analysis Key Takeaways}
Empathy plays a pivotal role in enhancing teamwork and fostering trust in AI recommendations within collaborative environments. The analysis reveals that empathetic communication strategies, particularly those emphasizing intent, can influence team dynamics. By promoting an environment where team members feel understood and valued, empathy facilitates open dialogue, which is crucial for effective collaboration. However, the variability in outcomes depending on the type of empathetic response underscores the need for strategic implementation. For instance, while neutral and agreeing responses may bolster trust, suggesting responses can inadvertently create tension. This complexity necessitates a careful calibration of empathetic expressions to align with team goals, thereby ensuring that AI trust remains robust, and that team cohesion is strengthened.

The socio-cognitive dimension emphasizes the importance of communication framing in influencing team performance and AI trust. By focusing on positive connotations and appropriate moral foundations, teams can create an environment conducive to collaboration and trust. The analysis highlights that positive framing not only enhances team coordination but also reinforces the reliability of AI recommendations. Conversely, poorly chosen moral cues can lead to misunderstandings and diminish trust, illustrating the critical need for thoughtful communication strategies. As teams increasingly rely on AI for decision-making, understanding the socio-cognitive underpinnings of communication can enhance interactions, improve trust dynamics, and ultimately elevate team performance.

Emotional dynamics within teams are crucial in shaping AI trust and overall collaboration. The findings indicate that fostering positive emotional states, such as joy, enhances both team performance and trust in AI recommendations. This suggests that an emotionally supportive environment can promote constructive interactions and bolster team effectiveness. In contrast, negative emotions, particularly sadness, can erode trust and hinder communication, leading to detrimental effects on teamwork. These insights highlight the necessity for leaders to actively cultivate a positive emotional climate, as emotional well-being not only facilitates better collaboration but also strengthens reliance on AI outputs. By prioritizing emotional health within teams, organizations can harness the full potential of both human collaboration and AI integration, leading to improved decision-making and performance outcomes.

\subsection{Preliminary Simulations of HDT Trust}
The causal relationships established by the previous finding allows for the use of these socio-cognitive-emotional measures as a proxy to judge a large language model’s capacity to believably simulate AI trust in future human digital twin simulations. Here we present a preliminary HDT experiments with the goal of investigating the capacities across large language models (LLMs) for generating simulated team communications from an assortment of closed and open-source models. 

Given the scope of this investigation, only team communications were modeled.
Without progress realistically tracked, simulation runs were arbitrarily terminated after a preset number of generations (100 turns). Communication dynamics such as conversational entropy and dominance were simulated by asking the model which HDT would like to speak next, and sequence length termination followed default settings for each LLM. 

A tiered prompting approach was used to test how well each LLM could replicate the kind of language used by real players from ASIST experiments. The base prompt included an overview of the mission itself, and a persona to describe the nature (i.e. capabilities and responsibilities) of each HDT’s role within the mission. The next condition was to also include example chat logs from a real mission where the players demonstrated above average success in completing objectives. Lastly, in the third condition, the outcomes of the mission were also included, which indicated that the provided logs were an exemplar and ought to be emulated for success. Note that explicit prescriptions of personality traits were not included in any experimental conditions.

A total of 5 models of different sizes and level of public availability were tested (gpt-4o-mini, gemma2\_9b, llama3.1\_8b, mistral\_7b, phi3\_14b) across the three conditions, 20 times each, for 100 turns per simulation. All provided logs were of the same exemplar mission, from which the socio-cognitive-emotional baseline outcomes were extracted, as seen in the following plots.

\begin{figure}
    \centering
    \includegraphics[width=0.49\linewidth]{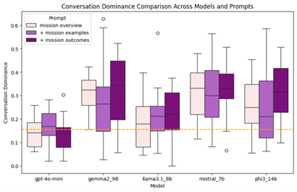}
    \hfill
    \includegraphics[width=0.49\linewidth]{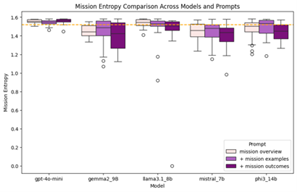}
    \caption{Deviations of simulated conversations from the ground truth (orange) across model and level of prompt  with respect to conversational dominance (left) and entropy (right).} 
    \label{fig:4-deviation-simconvo}
\end{figure}

\begin{figure}
    \centering
    \includegraphics[width=0.49\linewidth]{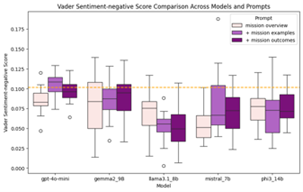}
    \hfill
    \includegraphics[width=0.49\linewidth]{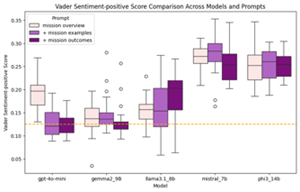}
    \caption{Deviations of simulated conversations from the ground truth (orange) across model and level of prompt with respect to negative (left) and positive (right) sentiment}
    \label{fig:5-deviation-simconvo}
\end{figure}

\begin{figure}
    \centering
    \includegraphics[width=0.49\linewidth]{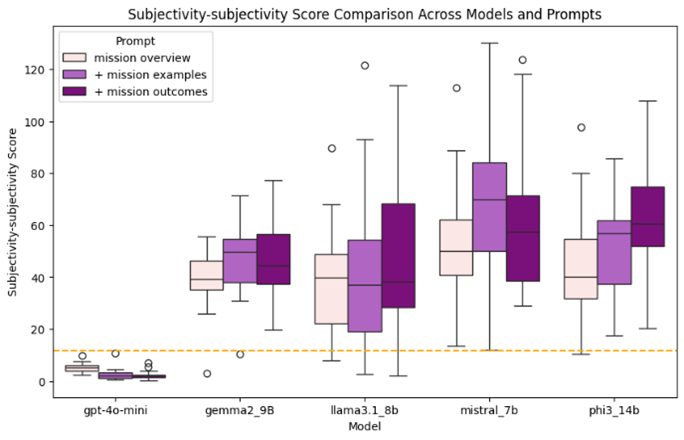}
    \hfill
    \includegraphics[width=0.49\linewidth]{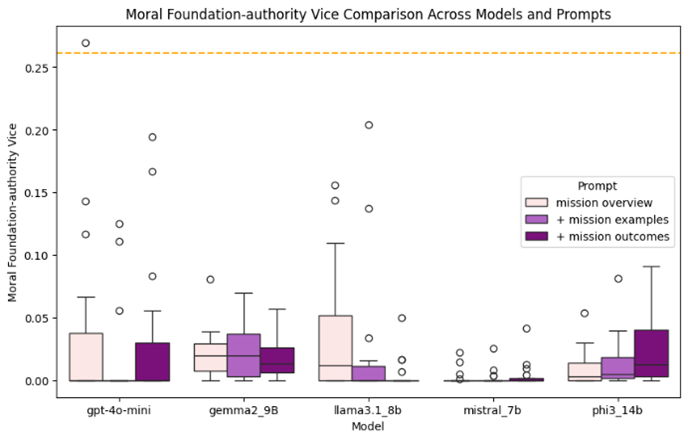}
    \caption{Deviations of simulated conversations from the ground truth (orange) across model and level of prompt with respect to the use of subjective language (left) and appeals against authority (right)}
    \label{fig:6-deviation-simconvo}
\end{figure}

The ability of each model to simulate realistic conversation dynamics was analyzed in terms of dominance, turn order entropy, and sequence length variance as shown in Figure~\ref{fig:4-deviation-simconvo}. We observed that the open-source model Llama 3.1 8b is most prone to misunderstanding the prompt and creating outliers. In one such example, HDTs would attempt to recreate the entire simulation themselves within a single reply. All models demonstrated a deviation from baseline human sentiment that skewed towards excessive positivity  as shown in Figure~\ref{fig:5-deviation-simconvo}. OpenAI’s GPT-4o-mini did the best at overcoming this bias when provided additional context on how the conversation actually unfolded. The use of subjective language across all models except GPT-4o-mini was much higher than the baseline, while the use of language indicating appeals against authority was relatively rare in all models  as shown in Figure~\ref{fig:6-deviation-simconvo}. 

\subsubsection{HDT Simulation Key Takeways and Limitations}
Our preliminary simulations revealed several important insights and limitations regarding the use of HDTs for modeling trust in human-AI teams. A fundamental challenge emerged in the form of baseline behavioral deviations from realistic human interactions. Despite using sophisticated language models and varied prompting approaches, the simulated interactions consistently demonstrated patterns that diverged from authentic human team communications. These deviations suggest that current HDT implementations, while promising, require more advanced prompting strategies and architectural considerations to achieve higher fidelity with human behavior patterns.

The simulation results highlighted a particularly interesting challenge regarding the representation of diverse personality types and interaction styles. Our findings suggest that current HDT implementations tend toward normalized, socially acceptable behaviors, making it difficult to model the full spectrum of human personality characteristics that might affect team dynamics. This limitation becomes especially relevant when considering the need to understand how different personality types—including more challenging or disruptive personalities—influence trust development and team performance in HAT contexts.

A critical direction for future research lies in exploring how HDTs can more accurately model diverse personality types, including those that might be considered toxic or extreme. This avenue of investigation is particularly valuable because studying such personalities in real-world HAT scenarios presents significant logistical and ethical challenges. However, this research direction faces its own set of obstacles. Commercial LLM providers have implemented robust safety measures in response to public scrutiny and concerns about model misuse, making it challenging to simulate more extreme personality types, even in controlled research contexts.

The tension between safety constraints and research needs is further complicated by the sophisticated methods that have been developed to circumvent these safety features, as documented by~\cite{shahScalableTransferableBlackBox2023}. While such bypass techniques exist, their use raises important ethical considerations about responsible AI research. This challenge points to a broader question about how to balance the need for comprehensive HAT research with responsible AI development practices.
\section{Research Question 2: What characteristics of trust must be operationalized in HDT trust models?}

To inform the most important components of HAT trust for HDTs to model, we begin by reviewing three major considerations of HAT trust that must be incorporated in concert to holistically model HAT trust. Key to understanding trust is its longitudinal nature, which is not only initially set, but also calibrated over time \cite{chiouTrustingAutomationDesigning2023, devisserTheoryLongitudinalTrust2020}. That is, an initial trust intercept is set, from which trust increases and decreases in response to trust violation and repair. To this end, modeling HAT trust requires approximating 1) how initial trust values are set, 2) the rate of changes in trust, and subsequently 3) operationalizing the time-dependent aspects of trust accordingly. 

\subsection{Setting HDT Initial Trust in AI Teammates}
First, regarding initial trust, several dispositional characteristics have been noted to be of key influence in setting a human’s initial trust of their agents \cite{merrittNotAllTrust2008}. Individual differences in trust tendencies refer to the innate or baseline levels of trust that individuals exhibit toward others, including AI-agent teammates. These tendencies shape how trust is initially formed and can impact the dynamics of human-agent interactions within a team setting. Of note, propensity to trust is a widely recognized component of a person’s general willingness to rely on others, including non-human entities. Jessup et al. \cite{jessupMeasurementPropensityTrust2019a} emphasize that individuals with a high propensity to trust are more likely to extend trust to AI agents, even in early stages of interaction. This tendency is often influenced by past experiences, cultural backgrounds, and personality traits \cite{merrittNotAllTrust2008}. In designing HDTs, it is crucial to incorporate these individual differences in trust tendencies to achieve a more accurate and personalized representation of initial trust. HDTs should be capable of simulating how a particular human might respond to various AI agent behaviors based on their innate propensity to trust. This could involve calibrating the HDT's initial trust levels to reflect the human's baseline trust tendencies and adjusting these levels as the simulated interactions progress. Such an approach enables the HDT to model how trust might evolve uniquely for different individuals, thereby providing more nuanced insights into human-agent teaming scenarios.

\subsection{Changes in Trust Over Time}
Second, regarding longitudinal changes in trust, both characteristics of the human and characteristics of the event influence the magnitude of trust increases and decreases. Emergent trust patterns refer to how trust develops, changes, and adapts over time within human-agent teams. Trust in this context is not static; it evolves as the human and AI agent interact, especially in response to specific events such as trust violations and subsequent repair efforts. Trust violation and repair is a critical aspect of emergent trust patterns, where trust is compromised due to a perceived failure or breach by the AI agent. Wildman et al. \cite{wildmanTrustHumanAgentTeams2024} describe trust violation as a dynamic process influenced by factors such as the severity of the breach, the frequency of violations, and the perceived intentions of the AI agent. The process of trust repair, in turn, involves behaviors such as offering explanations, apologies, or demonstrating improved performance, which can help restore trust over time \cite{bakerUnderstandingTrustRepair2018b}.

To accurately model emergent trust patterns, HDTs need to incorporate mechanisms for simulating trust violation and repair processes which negatively effects AI and HAI safety \cite{yaghiniRegulationGamesTrustworthy2024}. This includes the ability to model a human’s attitudinal and behavioral trust reactions to an AI agent’s mistakes, and how that trust could potentially be rebuilt through subsequent interactions. An HDT that effectively captures these dynamics would be better equipped to predict how trust might evolve in real-world human-agent teams, providing valuable insights into designing AI systems capable of maintaining and restoring trust over time.

\subsection{Computational Modeling of HDT Trust Over Time}
Third, modeling longitudinal trust values must be measurable. That is, approximations of human trust via the use of a HDT requires that the values and form of these trust trajectories are correspondingly output and analyzed. The measurement of trust characteristics is essential for understanding how trust develops, fluctuates, and stabilizes within human-agent teaming environments. Various quantitative and qualitative approaches have been employed to measure trust levels and their evolution over time.

Growth modeling is one such method that offers insights into how trust changes throughout the lifespan of a human-agent collaboration. Abramov et al. \cite{abramovParadoxicalDeclineGrowth2020} highlight the importance of using longitudinal growth modeling techniques to capture the trajectory of trust development, accounting for both linear and non-linear changes. By modeling trust as a dynamic construct, researchers can identify critical periods when trust is most susceptible to influence and intervention. When operationalizing trust within HDTs, it is vital to integrate growth modeling techniques that allow for tracking and predicting trust evolution across different stages of interaction. An HDT should be able to simulate not just a static level of trust but also how trust characteristics might shift in response to various situational factors, such as the AI agent’s performance consistency, transparency, and adaptability. This ability to model trust growth and decay over time ensures that the HDT provides a realistic and comprehensive representation of human-agent trust dynamics.

\section{Research Question 3: How do experimental manipulations from HAT studies translate to HDT studies?}
Although modeling trust dynamics in HDTs offers a viable alternative to human-subject studies for approximating HAT trust, not all experimental manipulations effectively translate to HDTs. Manipulations from the HAT literature may not always bear direct one-to-one approximation for reasons related to an HDT’s inability to simulate complex human phenomenon (e.g., emotions, physiology, sensation, and perception). In this section, we first expand upon these reasons that HAT trust manipulations may not maintain their validity in HDT studies, then share our initial foray into studying the HAT trust manipulations that do maintain their validity and provide meaningful insight into the validity of HDT trust.

\subsection{Replicability of HAT Trust Manipulations}
In order to even measure and capture a HDT’s reactions to AI agents, the HDT must first possess the ability to appropriately perceive and react to any manipulations in question. While HDTs aim to replicate human behaviors and cognitive patterns, the nuances of human emotion, cognition, and social interaction may not always be perfectly mirrored in a digital twin. In addition to the measurement concerns discussed in the sections above, this difficulty in accurately capturing these nuances also poses issues for designing experimental paradigms involving HDTs. 

At the core of this discussion lies a key question: what human characteristics are not replicable by a HDT? While HDTs can be designed to mimic known human patterns, they face difficulty in mimicking the underlying reason behind the patterns. Otherwise put, HDTs can replicate observable behaviors and thinking patterns, but cannot replicate the complex mechanisms that drive these observable behaviors and patterns. Key examples of these irreplicable mechanisms include emotions (i.e., when and how discrete affective states form), physiology (i.e., physical human biology responsible for human function), and sensation/perception (i.e., how humans holistically register information from their five sense and make meaning out of the sensory input). 

Because HDTs are unable to replicate these complex human mechanisms, several HAT trust manipulations that critically depend on these mechanisms for driving human reactions do not translate well for HDT studies. Examples of HAT trust manipulations that lose validity in HDT studies due to emotion-related limitations include manipulations of the AI agent’s integrity and affectively charged trust violations. Because these manipulations are driven by affective and attitudinal processes to influence trust ratings \cite{alarconAffectiveResponsesTrust2024}, HDT’s likely do not have the same reaction to these manipulations as a human would in terms of both magnitude (i.e., how much trust is loss from the moment the manipulation is registered) and longevity (i.e., how long the manipulation continues to impact trust after the moment the manipulation is registered). 

Examples of HAT trust manipulations that lose validity in HDT studies due to physiology-related and sensation/perception-related limitations include embodiment and anthropomorphism. Because an AI agent’s embodiment refers to the manifestation of the AI agent (i.e., virtually through an avatar on a screen, physically as a robot, or invisible as an algorithm; \cite{gliksonHumanTrustArtificial2020a}), which is registered through visual, auditory, or tactile senses, HDTs do not perceive or make meaning of the agent’s embodiment in the same way a human does. Similarly, because anthropomorphism depends on a human’s perception of human-likeness in an AI agent to drive reactions \cite{crowellAnthropomorphismRobotsStudy2019}, manipulating the AI agent’s anthropomorphic qualities is unlikely to produce similar sense-making and perception processes in a HDT as a human. 

While the above manipulations may not effectively translate to HDT studies of HAT trust because they depend on these complex human mechanisms to primarily impact changes in trust, several HAT trust manipulations are intended to impact trends in trust or trust behaviors. In light of this, there are still many promising avenues for experimentally studying HAT trust using HDTs. 

\subsection{Designing a HDT Trust Experiment}
Several manipulations of HAT trust still translate well to HDT studies, with perhaps some of the most insightful manipulations centering around the key characteristics of trust discussed in research question 2. Fundamentally, the core question surrounding the validity of HDTs for approximating trust is whether or not it can effectively replicate human trust patterns and behaviors. In light of this, experimental manipulations that are known to impact a human’s trust level at multiple points in time (i.e., initial trust and event-based changes in trust over time; \cite{bakerUnderstandingTrustRepair2018b}) or a human’s trust behaviors (i.e., use, misuse, or disuse; \cite{parasuramanHumansAutomationUse1997}) are likely to yield meaningful reactions in a HDT and produce insightful approximations of human trust. 

To this end, in our upcoming validation study, we designed a 3 (HDT manipulation) x 2 (AI agent manipulation) experimental framework that focuses on manipulations that have been known to impact trust at multiple points in time as well as impact trust behaviors. Regarding HDT manipulations, we focus on dispositional manipulations have been found to play an important role in setting a human’s initial trust in an AI agent as well as the malleability of trust building over time. Specifically, we manipulate the HDT’s levels of openness \cite{aliasghariEffectDomesticTrainee2021, zhangAutomatedVehicleAcceptance2020}, agreeableness \cite{bawackExploringRolePersonality2021, lyonsRoleIndividualDifferences2020}, and propensity to trust AI \cite{jessupMeasurementPropensityTrust2019a}. Because this study is a product of an effort meant to develop and validate HDTs, we maintain parsimony and set each of these manipulations to have two levels, either high or low. Similarly to how we would expect a human to react based on extant literature, we predict that when the HDT has high levels in each of these variables, their initial trust levels will be higher than when the HDT has low levels of the respective manipulations. Additionally, we expect that when the HDT has high levels of these manipulations that building trust will be more malleable over time. That is, it will be easier for the HDT to increase their trust in the AI agent over time compared to when they are set to low levels of these manipulations. 

Regarding manipulations of the AI agent, we manipulate the AI agent’s transparency and competency, with both manipulations having two levels (high and low). Manipulations of transparency refer to how clear and understandable the AI agent’s decision-making process is perceived to be by the HDT \cite{mercadoIntelligentAgentTransparency2016}. Within the HAT trust literature, the AI agent’s transparency has been observed to increase a human’s trust in the AI agent  \cite{chenEffectsAgentTransparency2016}. In the high transparency condition, the AI agent states the decision it intends to make along with the information and reasoning that led to its decision whereas in the low transparency condition, the AI agent simply states its decision without any explanations. In line with the HAT literature, we expect that the HDT will lose less trust in the high transparency condition than the low transparency condition. 

Manipulations of competency refer to how effective the AI agent is when completing tasks and fulfilling its role within the team \cite{fanInfluenceAgentReliability2008}. Within the HAT literature, when AI agents are perceived as competent (e.g., performing above a certain threshold; \cite{yuUserTrustDynamics2017}), a human team member will increase their trust in the agent over time. Conversely, if the AI agent is perceived as insufficiently competent (e.g., performing below a certain threshold; \cite{yuUserTrustDynamics2017}), a human team member will lose more trust in the agent over time.  trust levels in the HDT may decrease, particularly if the HDT already has a low baseline trust level. In the high competency condition, the AI agent consistently performs their tasks correctly on time, whereas in the low competency condition, the AI agent makes mistakes when performing their task by either providing inaccurate information or failing to complete their task. In line with the HAT literature, we expect that the HDT will gain trust over time in the high competency condition but lose trust over time in the low competency condition. 

By investigating the intersection of HDTs and trust within HATs, this research aims to advance the discourse on how AI agents can be designed and deployed in ways that foster effective collaboration and mutual understanding, ultimately leading to improved team performance and outcomes. This paper not only highlights the critical role of trust in human-agent teaming but also positions HDTs as a transformative tool for modeling and analyzing trust dynamics in this rapidly growing field.

\section{Conclusion}
This paper has explored the potential of human digital twins (HDTs) for modeling trust in human-agent teams, addressing key questions about the operationalization, measurement, and experimental manipulation of trust in HDT contexts for operational scenarios. Our investigation reveals both promising avenues and important challenges in using HDTs to advance our understanding of trust dynamics in HATs.

\subsection{Contributions to HAT Science}
The integration of HDTs into HAT research offers contributions to the field. Firstly, our exploration provides valuable insights into how LLM-powered HDTs can be effectively used to simulate and study trust behaviors in human-agent teams. By modeling key aspects of human trust tendencies, including individual differences and emergent trust patterns, HDTs offer a novel approach to examining trust dynamics that may be difficult or resource-intensive to study with human subjects.

Secondly, this work opens new avenues for enhancing trust modeling and measurement in HAT systems. The use of computational models like HDTs allows for more precise control over variables and enables the exploration of a wider range of scenarios than might be practical in traditional human-subjects research. This approach can lead to more nuanced and comprehensive models of trust in HATs, potentially informing the design of AI agents that can better adapt to human trust tendencies.

\subsection{Future Directions}
While our research demonstrates the potential of HDTs in HAT trust research, it also highlights areas requiring further investigation. Future work should focus on refining trust measures in HDTs to ensure they accurately reflect the complex nature of human trust. This may involve developing new measurements that capture both the cognitive and emotional aspects of trust, as well as validating these measures against ground truth data.

Additionally, there is potential in exploring additional behavioral and cognitive traits that HDTs could model in HAT settings. Beyond the trust-related factors examined in this study, future research could incorporate a broader range of human characteristics, such as risk tolerance, cultural background, or prior experiences with technology. This expansion would allow for a more comprehensive simulation of human diversity in HAT contexts.

Furthermore, longitudinal studies using HDTs could provide valuable insights into trust development and maintenance over extended periods of human-agent collaboration. Such studies could reveal patterns in trust dynamics that might not be apparent in shorter-term experiments.

In conclusion, while HDTs offer exciting possibilities for advancing HAT trust research, their effective use requires careful consideration of their current limitations and ongoing validation against ground truth data. As we continue to refine HDT models, we can expect to gain deeper insights into the nuances of trust in human-agent teams, ultimately contributing to the development of more effective and harmonious collaborations between humans and AI agents.

\section{Acknowledgments}
This work is supported by the Defense Advanced Research
Projects Agency (DARPA) contracts HR00112490410 and HR00112490408.
The views, opinions, and/or findings expressed are those of
the author(s) and should not be interpreted as representing
the official views or policies of the Department of Defense or
the U.S. Government.

\bibliography{Nguyen-etal-HumanDigitalTwinTrust}

\newpage
\section*{Biographical Note}
Dr. Daniel Nguyen is an Associate Research Scientist at Aptima, Inc. His research interests and experiences are focused on work teams, with a focus on human-agent teaming (HAT) which has led him to a broader interest in related human-factors topics such as human-performance and trust in automation. He received his M.S. and Ph.D. in Industrial/Organizational Psychology at Florida Institute of Technology in 2020 and 2023, and his B.A. in Psychology at Texas A\&M University in 2017.

Myke C. Cohen is an Associate Research Engineer at Aptima, Inc, and a Ph.D. student in Human Systems Engineering and CHART Scholar at Arizona State University. His research focuses on the design, modeling, and assessment of human-machine interactions in safety-critical work systems. He holds a Master of Science in Human Systems Engineering from Arizona State University and a Bachelor of Science in Industrial Engineering from the University of the Philippines Diliman.

Hsien-Te Kao is an Associate Research Engineer at Aptima, Inc., and a Ph.D. student in Computer Science at the University of Southern California. His research interests include cold start prediction, personalization, behavioral modeling, decision-making, and mobile health. He holds a Bachelor of Science in Mathematics from California State Polytechnic University, Pomona.

Mr. Grant Engberson, Associate Research Engineer, Aptima, Inc., specializes in artificial intelligence, neuroscience, and brain-machine integration. Expertise in neurological pathophysiology provides him with a unique perspective on machine learning, particularly at the interface of human and machine intelligence. He is skilled in biosignal processing, computer vision, natural language processing, rapid prototyping, and cortical surgery. Mr. Engberson received an MS in biomedical engineering from Northwestern University, and a BS in chemical and biochemical engineering from the Colorado School of Mines.

Mr. Louis Penafiel, Research Engineer and Lead, Aptima, Inc., works with the Artificial Intelligence Technologies capability—a portfolio of research projects focused on modeling and understanding natural language discourse, patterns of life, and planning. Prior to Aptima, he conducted data science research at CERN’s Large Hadron Collider and NASA’s Jet Propulsion Laboratory. Mr. Penafiel holds an MS in physics from Cornell University and a BS in physics and mathematics from the University of California, Riverside.

Spencer Lynch is Senior Software Engineer at Aptima, Inc., and specializes in high-performance, web-based applications and full-stack development, utilizing the latest web technologies such as WebGL, WebVR, WebRTC, and WebSockets. He focuses on developing high-quality AR and VR immersive simulations for training in various domains, such as spatial disorientation events during piloting, maintenance training, and location finding of hazardous radioactive material. Mr. Lynch is a member of SIGGRAPH and holds a BS in computer science from Bowling Green State University.

Dr. Svitlana Volkova is Chief of AI, Office of Science and Technology, Aptima, Inc., and a recognized leader in the field of human-centered artificial intelligence (AI). She leads the development of AI-powered descriptive, predictive, and prescriptive decision intelligence and analytics to model and explain complex systems and behaviors to address national security challenges in the human domain and beyond. She serves as senior PC member and area chair for top-tier AI conferences and journals including AAAI, WWW, NeurPS, ACL, EMNLP, ICWSM, PNAS, and Science Advances and is a senior board member for Women in Machine Learning. She received her PhD in computer science from The Johns Hopkins University, where she was affiliated with the Center for Language and Speech Processing and the Human Language Technology Center of Excellence.

\end{document}